\documentstyle[11pt]{article}
\def\rf{\hfill\break}
\def\etal{{\it et al. }}

\def\pt{\ \ \ \ .}

\def\sm{M_{\odot}}

\def\e{\eta}
\def\ep{\epsilon}
\def\g{\gamma}

\def\a{\alpha}
\def\b{\beta}

\def\p{\partial}

\def\D{\Delta}

\def\st{\sigma _T}
\def\ln{{\rm ln \ }}

\def\sles{\lower2pt\hbox{$\buildrel {\scriptstyle <}
   \over {\scriptstyle\sim} \ $}}
\def\sgreat{\lower2pt\hbox{$\buildrel {\scriptstyle >}
   \over {\scriptstyle\sim} \ $}}
\def\pac{Pac\'zynski }
\def\MR{M\'es\'zaros and Rees }
\def\RM{Rees and M\'es\'zaros }
\begin{document}

\centerline {\bf GAMMA RAY BURSTS FROM INTERACTION OF RELATIVISTIC}
\centerline {\bf FLOWS WITH RADIATION FIELDS}
\bigskip
\bigskip

\centerline{\bf Amotz Shemi}
\bigskip
\centerline{ Wise Observatory \& School of Physics and Astronomy}
\centerline{ Raymond and Beverly Sackler Faculty of Exact Sciences}
\centerline{ Tel Aviv University, Tel Aviv, 69978, Israel.}

\bigskip
\bigskip
\rf
{\it Accepted for publication in the MNRAS, April 1994}

\begin{abstract}
Relativistic flows resulting from sudden explosive events
upscatter ambient interstellar photons of local
radiation fields. For Lorentz factor $ > 100$ and
dense optical - UV radiation fields
the emergent signal is a typical
gamma ray burst. Presumably the explosions occur
in dense globular clusters or in galactic nuclei, at cosmological distances.

\end{abstract}

\rf

\section { Introduction}

The nonthermal spectra of gamma ray bursts (GRBs) pose
a generic problem for optically thick models of these bursts.
Independent of the details of a specific model,
the bursts, whether originating at cosmological distances or nearby
(but further than $ \sim 2$ pc),
are extremely compact and thus
inevitably involve enormous
pair-production
(Cavallo \& Rees 1978, Piran \& Shemi 1993).
They
are therefore optically thick even if they are free of baryon contamination.

A pure hot pair-plasma fireball (Goodman 1986) of initial temperature
$T_0 $ and
dimension $R_0$ expands and cools
until it extends to $\sim R_0 \times (T_0/25 {\rm KeV})$,
where the pair - production optical depth falls below unity.
The observed spectrum
of the blue shifted photons is a modified blackbody at the initial temperature,
in contrast to GRBs.
In the later case the spectrum is a rather smooth
power-law, sometimes shows a high energy tail above 1 GeV, and does not peak in
the vicinity of the
0.51 MeV annihilation line
(some bursts show a broad bump at $\sim 430$
KeV that can be interpreted as a red-shifted 0.51 MeV line).
The conclusion is that pairs annihilate
at some optically thick
phases and that
additional mechanisms are involved and become dominant at later stages.

More realistically, the fireball is contaminated by baryonic matter, whether
released
during the explosion or
surrounding the explosive object beforehand.
This load of matter poses another severe prenhfireball opacity and greatly
affects its
dynamics.
Even for a small contamination of $\sim 10^{-7} \sm $
a large fraction of the available fireball energy is converted
into kinetic energy of the outflow. The flow is accelerated
to ultrarelativistic velocities and the final
signal is weakened and
shifted to the UV range (\pac 1990 , Shemi \& Piran 1990).

Different cosmological
scenarios, e.g. binary neutron star mergers, black hole - neutron star
mergers,
normal stars disrupted by a giant
black hole, accretion - induced collapse of white dwarfs to neutron
stars or the collapse
of massive stellar cores,
share a general scenario in which
$\sim 10^{53}$ ergs are suddenly injected
via a burst of neutrino-antineutrino emission.
Matter, possibly a few
percent of a solar mass, is expelled
by the burst and more of which
envelops the central object, sometimes in a configuration of a (non-stable)
disk (e.g. Woosley 1993).
Almost all of the explosion energy is
dissipated by the
neutrinos and gravitational waves, but a
small fraction of it creates an optically thick pair-plasma fireball
of $E_0 \sim 10^{51}$ erg, that expands, cools and
becomes diluted.

The fireball
is rearranged into a configuration of a thin shell,
of width $ \Delta  r $ that roughly equals the initial radius
$R_0 \ \sgreat \  10^6$ cm,
and expands ultra-relativistically with  $\g \sim \e \equiv E/Mc^2$
(Piran Shemi \& Narayan 1993, hereafter PSN93).
Photons would escape only when the gas opacity falls,
long after the temperature has been degraded below the pair production
threshold and the pair annihilation was completed, with
a spectrum of modified blackbody (Shemi 1993a).
\pac (1990) Discusses a steady state
ultrarelativistic super-Eddington wind, rather than an instantaneous burst, and
finds results qualitatively similar to those of the transientdescribed above.

Both the
nonthermal spectrum and the
high sensitivity to baryon overload
are explained if
the actual burst is
emitted at later stages, when additional mechanisms
reconvert the bulk kinetic energy to radiation
(\RM 1992, hereafter RM92).
Particularly, RM92 suggest deceleration of the
fireball (or the relativistic 'wind')
by the ambient interstellar matter, which is swept up in a relativistic shock
wave, similarly to supernovae remnants, but
typically of a $\g$-ray signal (see also \MR 1993, hereafter MR93,
Katz 1994). A burst of $\g$-rays emerges when the compressed matter in the
blast wave shell cools. This model was strongly criticized by
Colgate (privet communication) because of
the intense $e^+e^-$ pair production in the shocked matter that would
enhance the opacity and reduce the radiation energy.
The expected output of the shock would basically be accelerated
matter, like cosmic rays (Colgate 1975), rather then a strong $\g$-ray pulse.

In this paper I suggest an alternative
mechanism for energy reconversion in cosmological GRBs.
An ultrarelativistic wind interacts with
a local thermal radiation field,
presumably of optical photons (Shemi 1993b).
The electrons upscatter the ambient photons,
shift them towards the flow direction and modify
the spectrum to a power law.
The rather long duration of GRBs
and their rapid time variability are
naturally solved if the radiation field extends over a rather large
scale of $\sles 10^7$ light seconds and is inhomogeneous.
The relative importance of the
braking by inverse Compton process and by external matter depends on the
ratio of the
mass-energy densities in the radiation and in the external matter.

\section {A Scenario}

Consider an ultrarelativistic wind ($\g \gg 1 $)
of ionized matter (electrons and protons) that
enter a radiation field of low temperature (in the optical - UV rangeinverse
Compton (IC) process, raise their energies
by $\sim (4/3) \g^2$ and shift them
towards the flow direction.
For example, Lorentz factor $\g \sim 380$ is sufficient
to boost 2.5 eV photons to 500
KeV $\gamma$-rays.

The power-law spectra of GRBs requires complex
wind, rather than monoenergetic, of broad energy distribution.
A single shell resulting from a homogeneous fireball
is basically monoenergetic,
but realistic cosmological explosions with prolific
$\nu \bar \nu \rightarrow e^+ e^-$ pair production, such as NS-NS mergers or
BH-NS mergers,
are inhomogeneous (e.g. RM92). Also, the baryon load would spread nonuniformly,
concentrated in the orbital plane in case of binary a system.
The overall result is
a mix of Rayleigh-Taylor unstable shells, with varied $\g$ factors.
For example, if a fireball front propagates down a radial density gradient the
distribution is
roughly a power law $ \propto \g^{-\alpha} $ with $\alpha \sim 1/2$ (e.g.
Colgate 1975).

If the ambient radiation field contains significant amount of
high energy photons
(e.g. UV and X-ray photons from X-ray sources and hot stellar coronae in
dense cores of globular clusters)
it would also enhance the high energy tail of the final spectrum.
Also, if the field is nonuniform, the flow, actually, will passe several fields
of different temperature and photon density.
The time - integrated spectrum would be a blend of
thermal like spectra, emitted subsequently from different regions
on extremely short receiving time intervals.

Almost all the flow kinetic energy is associated with the baryons, but it would
transferred to the photon field via electrons.
The electrons and
the baryons are assumed to be coupled by electrostatic forces, thereby
the kinetic energy would if each electron scatters
$$
2.8\times 10^5 \g_3^{-1}\ep_{2.5eV}^{-1}
\eqno(1)
$$
times with
the radiation field, before it is decelerated by another mechanism.
Here $\ep_{2.5eV}= \ep/ 2.5$eV
is the mean
photon energy and $\g_3 \equiv \g /10^3$.

Let us recall some relevant distance scales in cosmic fireball hydrodynamics
(SP90, MR93, PSN93).
The source dimension is usually derived from burst rising time and
time variability $\Delta t /c \sim $ a few $\times 10^6 R_6$ cm.
The saturation radius defines the scale where the acceleration
saturates, and is
related to the maximal average value of
$\g  \sim \e $: $r_{sat} \sim 10^9 \eta_3
R_6$ cm ($\e_3 = \e/10^3 $).
When the expanding shell
reaches distance
$r_t \sim 1.9 \times 10^{13}\a^{-1} E^{1/2} \eta_3^{-1/2} $ cm
from the origin
its electron density falls
and it becomes optically thin to Thomson scattering
(here the opening half angle $\a$ stands for the case of beaming).
Finally,
unless substantial deceleration takes place beforehand,
the interstellar matter drags the flow and
hydrodynamical shocks are likely to be formed at
$r_{ISM} \sim 10^{16} E_{51}^{1/3} n_1^{-1/3}\g_3^{-2/3} $cm,
where $n$ is the mean ISM density, in units of one particle per ${\rm cm}^3$.

To obtain efficient IC deceleration, compared with the ISM drag, one needs
ambient radiation
fields of energy density larger than the rest mass density, which are
located relatively close to the explosive object, namely at
$ r < r_{ISM}$. In optimal conditions, unlike ISM drag, the IC efficiency
can be as large as $100\%$. Initially,
the radiation field energy density is negligible compared with
that in the onset phase of the fireball, therefore I neglect cases where
$r < r_{sat}$.
The transition
to an optically thin fireball occurs somewhere
between $r_{sat} < r_t < r_{ISM}$. In the following I will
distinguish betweethe optically thick case and optically thin case.

If the optical depth is large ($r < r_t$)
the ambient photons, initially of random
directions, associate with the flow.
These photons contribute only negligible energy to the wind but
they substantially increase the photon number density.
Net energy is transferred to the radiation
while the wind temperature remains relatively low.
The flow is decelerated and the spectrum of the diffusing photons
is modified by pure dynamical effects, in addition to being affected by the
physical conditions in the photosphere as
in static (or Newtonian $v \ll c$) cases.
Modification of a
spectrum from thermal to power law, that occurs
in the photosphere of relativistic flows, is discussed in the appendix.
Nobili, Turolla and Zampiri
(1993, NTZ93) discuss somewhat similar phenomena in accreting neutron stars
emitting nonthermal X-ray radiation.
Following them I focus on
special-relativistic $\g \gg 1$ flows, and derive
analytically a spectral index equal to $1.34$.
This example
shows that a nonthermal spectrum, generally a power law, can be obtained.
Evidently,
the subject needs to be studied more deeply.

In the optically thin case, where the IC braking occurs at $r >r_t $, the
emergent
intensity is sensitive to the initial energy spectrum of the
relativistic electrons. If they are initially monoenergetic, the
time integrated intensity rises as
$\ep^{1/2}$ below $\ep_c \simeq 2 \g^2
kT/m_ec^2$ and falls exponentially $\propto \ep^{1/2} {\rm exp}(-\ep /\ep_c)$
for $\ep_c \ll \ep$ (
Zdziarski, Svensson and \pac 1991, hereafter ZSP91).
The global behavior of the spectra
is a blackbody, broadened in the low and the high energy sides.
To overcome the discrepancy from exponential decline to power law
we may consideIC spectrum that is corresponding to
a power law electron energy
distribution of spectral index $is also a power law, of spectral index $(p -
1)/2$. Also,
even when the
(comoving) electron distribution is thermal, the emergent intensity
is a power law of index $ -{\rm ln}\tau/{\rm ln}A$, where $A \sim 4/3\g^2$
is the photon energy amplification factor
per scattering,
$\tau = \st n_e \D r $ is the optical depth to Thomson scattering,
and $\st $ is the Thomson cross section
(Rybicki \& Lightman 1979).
Total energy amplification of soft photons is
important if
$A\tau \  \sgreat \ 1$, a plausible condition in our case.

Dynamic effects are also important
when considering the
differences between the time interval in the electron (comoving)
reference frame
$dt_{com}$, the local rest frame emission time interval $dt_{em}$
and the receiving time interval $dt_{rec}$
in the remote observer frame.
These intervals are related as (Rybicki \& Lightman 1979)
$$
dt_{rec} =
[1  - \b {\rm cos} (\theta )] dt_{em}
= \g [1  - \b {\rm cos} (\theta )] dt_{com} \pt
\eqno(2)
$$
Particularly, we are interested in the case where $\theta \ll 1
$ and $ \g \gg 1$,
especially when $\theta \sim 1/\g$.
 From Eq. 2 we have
$$
dt_{rec} \simeq { 1 + \g^2 \theta^2 \over 2 \g^2 }dt_{em}
\to {1 \over  \g^2} dt_{em}  \ ,
\eqno(3)
$$
namely the receiving time
interval in GRBs  can be shorter
than the emission time interval by a factor of $\g^2$, say $ \sgreat 10^{5}$
The typical observed burst longevity
of 10 - 100 seconds indicates that
the interaction region can extend to $\sles 10^{17}$ cm.
It is encouraging to note that this length scale allows substantial
IC braking before the ISM drag becomes important, provided the ambient
radiation field is sufficiently dense.

\section {Discussion}

The sourcis presumably a dense population
of stars.
The time variability of GRBs is explained if the
expanding fireball encounters radiation
fields of different temperature,of photon bulk flow, and variable photon
density. The
total temporal affect is a variable flux.
The observed softening in the burst spectral evolution
is explained by the deceleration of the
fireball front.
Specifically, such populations may occur in
dense cores of globular clusters.
For example, the core radius of M15 is
$\sim 0.08$ pc and the stellar population is a few $\times 10^5$.

The Compton drag is significant when the mean free path of an
electron in a random photon field,
$(\st n_{ph})^{-1}$, is smaller then the field dimension, $d$.
The efficiency of such braking is proportional to the
ratio $\zeta$ of number of interactions against the
total number $N$ required for complete braking of the electron (Eq. 1).
Such efficiency can be large
if $\tau_e \equiv \st n_{ph} d
\sgreat N^{1/2} = 531 (\g_3\ep_{2.5eV})^{-1/2}$.
Using $\zeta = [(\tau_e /531)(\g_3\ep_{2.5eV})^{-1/2})]^2$ and impose
$\tau_e \leq 531$,
we obtain,
for
$d = 0.1$pc ($\equiv d_{0.1pc}$) and $\zeta = 0.1$, the photon density
$n_{ph} = 8.5\times 10^{8} d_{0.1pc}^{-1}$photons cm$^{-3}$.
This braking dominates the braking expected by ISM
if the radiation energy density exceeds that of the matter,
namely if $n_{ph} > n_{ISM} \times m_p c^2/\ep
= 3.76 \times 10^8 \ n_1 \ep_{2.5eV}^{-1}$.

For practical purposes let us consider
the density level of $10^8 \equiv n_8$ photon cm$^{-3}$
sufficient for
significant Compton drag. This is an extremely large density,
larger by $\sim 17$ orders of
magnitude from the mean galactic, and even
the mean density in dense cores, $\sim N_* L /( 4 \pi d^2 c \ep ) = 2.8 \times
10^3
 N_{*,5}l d_{0.1}(where the stellar total number is $N_{*,5} = N_*/10^5 $ and
$ l_{\odot} = L/L_{\odot}$), is much smaller than it. However,
the photon density of a radiation field surrounding a single star
$n_{ph}(r)\simeq L/(\ep c 4\pi r^2 =
5.4 \times 10^{11} l_{\odot} r_{\odot}^{-2} \ep_{2.5eV}^{-1}$,
($r_{\odot} = R/R_{\odot}$).
For $n_8 = 1$ we have an effective radius $r_{eff, \odot} =  73 (230)$, if
$l_{\odot} = 1 \ (10)$.
Interaction with that field reduces the fireball energy only by
a small fraction, since the typical value of $\tau_e$ for a single
star is less than unity.
When the wind confronts the near vicinity of a star
its corona and stellar wind also affect the energy deposition, where
free-free scatterings, IC of corona X-ray photons and hydrodynamical shock
heating
can take place.
X-ray sources, such as low mass X-ray binaries,
were detected in 10 GCs (e.g. the LMXB source AC211 locates within 0.4 pc of
the core center of M15) with
typical luminosity either $3\times 10^{35}$ or $3\times 10^{37}$ ergs s$^{-1}$.
Ignoring these possibilities,
we note that an efficient drag requires $\tau_e$ of a few hundreds,
namely a case where the fireball encounters few hundred stars (of effective
radius
$r_{eff} \sim 10^2 R_{\odot}$) before escaping the cluster.
Unfortunately this case is rather difficult to obtain, as the
number of interactions regarded to the {\it averages} values
$\sim N_{*} (r_{eff}/d)^2 $ is less then unity. However,
in the centers of
many cores the radial mass distribution seems to peak, roughly
as a power law. Post-core collapse models predict extremely
dense cores, and also
the density of the gas and the dust there are very low. Thereby it is
plausible to expect
optimal conditions in some dense cores.

Cosmic fireballs in GCs are expected favorably from mer(Piran, Narayn \& Shemi,
1991).
Among the four systems known today, one of them (PSR2127+11C) is located
in M15, and this system is expected to merge due to gravitational radiation
emission
$2\times 10^8$ years from now.
Phinney (1991) considered
neutron star with a companion normal star in a star - neutron star binary
(Phinney and Sigurdsson 1991).
The conservative estimated GC merger rate is
$3 \times 10^{-9} h^{-3} \ {\rm Mpc^{-3} \ yr^{-1} }$,
small by an order of magnitude to meet the CGRO-BATSE event rate.
Sigurdsson and Hernquist (1992) suggest
an additional mechanism for creating double pulsars in GCs.
Encounters between pairs of hard binaries (semimajor axis $a \ll 10$ AU),
each containing a neutron star and a main sequence star, yield  double neutron
star
systems where the two normal stars disrupted to form a common envelope around
them.
The final eccentricity is expected to be small but the system is that close so
that
the orbital decay time is $10^{8} - 10^{9}$ yr. Also, there are
growing evidences for very large populations of globular
clusters ($\sgreat 10^4$) in giant elliptical galaxies (e.g Harris 1991)
which were not considered explicitly by Phinny (1991). This
may raise the estimated merger rate in GCs,
but probably more than a single mechanism
to create cosmic fireballs in GCs is required to meet the total GRB rate.

Dense radiation fields also characterize the central region of galaxies.
In this case the fireball source can be connected with
the physics of the galactic nucleus.
Particularly, Carter (1992) suggests the creation of a cosmic fireball
by the violent disruption of a normal star
plunging sufficiently close to
a giant black hole in a galactic nucleus.

To conclude, Compton drag of relativistic debriexplains several generic
problems in optically thick GRBs, i.e.
their large compactness,
the nonthermal spectrum,
the sensitivity to baryonic load, the time-scales and the time variability.
However, a main difficulty is to obtain large radiation density in the
extended vicinity
object.
Optimal cases of radiation fields in very dense cores of globular
clusters or in galactic nuclei
probably meet this requirements. The explosive events can be
mergers of binaries of compact objects in the first case, and
star disruptions in the second one.

It should be mentioned that
Compton drag in GRBs was discussed in a different context by ZSP91
who studied the creation of cosmological
GRBs at  $z \gg 1$ from an electron-positron cloud of relativistic bulk
motion that upscatter the ambient cosmic background radiation. Also, after this
study was
completed I became aware of the work of
Epstein \etal (1993),
who propose GRBs from
Compton drag of relativistic flows, originated by explosive
events in AGNs.

\bigskip
\rf
I thank G. Drukier, I. Goldman, J. Katz, D. Maoz, M. J. Rees, S. Sigurdsson and
B. Yanny for discussions.
\bigskip
\bigskip
\rf
{\bf References}
\bigskip
\rf
Carter, B. 1992, ApJ (Letters) {\bf 391}, L67.
\rf
Cavallo, G., \& Rees, M.J. 1978, MNRAS, {\bf 183}, 359.
\rf
Colgate, S.A, 1974, in {\it Origin of Cosmic Raye}, ed. Osborne, J. L. \&
Wolfendale, A. W.,

NATO Study, Durham, England, Reidel Pub., 447.
\rf
Epstein, R.I., Fenimore, E.E, Leonard, P.J.T. \& Link,
 B. 1993, in {\it Texas/PASCOS 92:

Relativistic Astrophysics and Particle Cosmology}, eds. A. W.
Akerlof and

M. A.  Srednicki; {\it Annals of the NY Academy of  Science}, {\bf 688}, 565.
\rf
Goodman,J 1986, ApJ (Letters), {\bf 308}, L47.
\rf
Harris, G.W., 1991, Ann. Rev. Ast. Ast. {\bf 29}, 543.
\rf
M\'es\'zaros, P. \& Rees, M.\rf
Nobili, L. Turolla, R \& Zampieri, L. 1993, ApJ, {\bf 404}, 686.
\rf
Pac\'zynski,B. 1986, ApJ (Letters), {\bf 308}, L51.
\rf
Pac\'zynski,B. 1990, ApJ, {\bf 363}, 218.
\rf
Phinney,E.S 1991, ApJ (Letters), {\bf 380} L17.
\rf
Phinney,E.S. \& Sigurdsson, S. 1991, Nature, {\bf 349}, 220.
\rf
Piran, T., NarAl, 1991,

eds. W.S. Paciesas \& G.J. Fishman, AIP press, 149.
\rf
Piran, T. and Shemi, A. 1993, ApJ (Letters), {\bf 403}, L67.
\rf
Piran, T. Shemi, A. \& Narayan, R. 1993, MNRAS, {\bf 263}, 861.
\rf
Rees, M.J. \& M\'es\'zaros, P.  1992, MNRAS, {\bf 258}, 41.
\rf
Rybicki, G.B., \& Lightman, A.P. 1979, {\it Radiation Processes in
 Astrophysics}, W\&S.
\rf
Shemi, A. \& Piran, T. 1990, ApJ (Letters) {\bf 365}, L55.
\rf
Shemi, A. 1993a; 1993b, in {\it Gamma Ray Burst} workshop, ed. Fishman, G.,

Hurley, K \& Brainerd, J., Huntsville,
Oct. 20 - 22.
\rf
Sigurdsson, S. \& Hernquist, L. 1992, ApJ (Letters), {\bf 401}, L93.
\rf
Thorne, K. S. 1981, MNRAS, {\bf 194}, 439.
\rf
Woosley, S. E., 1993, ApJ, {\bf 405}, 273.
\rf
Zdziarski, A.A., Svensson, R. \& Pac\'zynski,B. 1991, ApJ {\bf 366}, 343.

\bigskip
\rf
\centerline{\bf  APPENDIX: Radiation diffusion in ultrarelativistic flows}
\bigskip

The spectrum of IC radiation from the photosphere of
a relativistic wind is discussed in this appendix under some assumptions.
For simplicity
we consider a (gravitation, magnetic) field-free case,
using a steady state approximation, and treating
the spherically symmetric flow of a perfect fluid,
assuming it is optically thick in
$r_{min}$ and becomes optically thin at $r_{max}$.
The matter and the radiation are coupled via the momentum and energy
radiation hydrodynamics equations (NTZ93)

$$
(P + \rho) {d \ln \g
\over d \ln r} + {d P \over d \ln r} + {r s_1 \over \g } = \eqno(A1)
$$

$$
{d \rho \over d \ln r} - (P + \rho )  {d \ln \rho_0 \over d \ln r} -
{r s_0 \over v\g } = 0 \ \ ,
\eqno(A2)
$$
and the continuity equation is given by
$$
{d \ln u \over d \ln r} +  {d \ln \rho_0 \over d \ln r} + 2 = 0 \pt
\eqno(A3)
$$
Here $\g \equiv (1 - v^2)^{-1/2} $, $u = \g v $ (with $c = 1$) and
$s_0, \ s_1$ are tThe radial profiles of the hydrodynamical variables are
obtained by integration
over Eqs. A1 - A3, provided the source terms are known.
In principle, these terms can be obtained by
integration over a set of frequency-dependent radiation transfer equations.
The radiation can be described by moments
$w_i$ of the radiation intensity (Thorne 1981).
Here we overcome these difficulties
by assuming a deceleration of the form
$$
\g(r) \propto r^{-\a } \pt
\eqno(A4)
$$
Using the relations
$d \ln u /d \ln r = v^{-2}  \ d \ln \g /d \ln r = - v^{-2} \a $,
the continuity equation reads
$$
{ d \ln \rho_0\over d\ln r} = -2 + {\a \over v^2} \pt
\eqno(A5)
$$
The interaction of the matter and the
radiation is approximated by Thomson scattering.
In the diffusion approximation ($\tau \gg 1$)
we assume $ s_0 = 0 $, $s_1 = -{\tau \over r} w_1 $, and
we also can neglect higher
moments $w_i \ i\geq 2$,
noting that the accuracy lost can be $\sles \ 15\%$ (NTZ93).
The first two equations for the radiation transfer read (Thorne 1981, NTZ93):

$$
- v {\p w_0  \over \p \ln r} +
(1+\a ) {\p w_1  \over \p \ln r}
+ ({2 v \over 3} - {\a \over 3v} ) {\p w_0 \over \p \ln \nu} =
$$
$$
= (2v + {\a \over v} ) w_0  + 2(\a - 1)w_1
\eqno(A6)
$$
and
$$
{1 \over 3}{\p w_0  \over \p \ln r}
- v {\p w_1  \over \p \ln r} +
({2 v \over 5} - {3 \a \over 5 v}) {\p w_1 \over \p \ln \nu} =
\bigl( {12 v\over 5} - {\tau \over \g } -  {3 \a \over 5 v} \bigl) w_1 \pt
\eqno(A7)
$$
Defining

$$
\p x \equiv \p \ln r \p y \equiv \p \ln \nu \ ,
\eqno(A8)
$$
we have

$$
d w_0 = {\p w_0 \over \p x} d {\rm x} + {\p w_0 \over \p y} d {\rm y}  \ \ ;
$$

$$
d w_1 = {\p w_1\over  \p x} d {\rm x} + {\p w_1 \over \p y} d {\rm y} \pt
\eqno(A9)
$$
Equations A6, A7 and A9 can be written in a matrix form:

\[ \left( \begin{array}{cccc}
  -v &    {2v \over 3} -{\a  {1\over 3} &0     & -v & {2 v \over 5} -{3 \a
\over 5 v} \\
  d{\rm x} & d{\rm y}& 0 & 0 \\
  0 & 0 &   d{\rm x} & d{\rm y}
\end{array} \right)
\left( \begin{array}{c}
{\p w_0 \over \p x} \\
{\p w_0 \over \p y} \\
{\p w_1 \over \p x} \\
{\p w_1 \over \p y}
\end{array}\right)
=
\]

\[ = \left( \begin{array}{c}
(2v + {\a \over v} ) w_0  + 2(\a - 1)w_1 \\
\bigl({12 v\over 5} - {\tau \over \g } -  {3 \a \over 5 v} \bigl) w_1\\
d w_0 \\
d w_1
\end{array}\right)
\begin{array}{c}
{\ \ \ \ \ \ \ \ \ \ \ \ \ \ \ \ \ \ \ \ \ (A10)}
\end{array}
\]

For an ultrarelativistic
flow we can substitute
$v = -1 $
($v$ in Eqs. A6-A7 is {\it negative}
for outflow). Upon equating the determinant in Eq. A10 to zero, we
obtain the characteristic equation:

$$
{d{\rm y} \over d {\rm x}} =
{11 \a - 8 \over 10} \pm
{1 \over 10} \sqrt{\Delta}
\eqno(A11)
$$
where
$$
\Delta =
24 + 24 \a + 81 \a^2 \pt
\eqno(A12)
$$

A comprehensive discussion must involve
integration of Eqs. A6 and A7, e.g. by
using the numerical scheme of
NTZ 1993.
Given the spatial conditions on
$w_i$ on the boundaries and the spectrum at $r_{min}$, the
problem can be treated as a
boundary value problem in $r$ and an initial value problem in $\nu$, and
it is convenient to note
that $\Delta$ is always positive, so that
equations A6, A7 are $hyperbolic$.
For the purpose of this
paper, we restrict our discussion to an analytical
behavior of the solutions, assuming
the radiation intensity $w_0$ and the flux $w_1$
are smooth, differ$(r_{min},r_{max};\nu_{min},\nu_{max})$.

The free parameter $\a$ must be chosen in a way that global requirements
from the matter -- radiation interaction
are met. To insure deceleration, we conclude from Eq. A4 that $\a$
should be positive. From Eq. A5 we see that $\a \leq 2$, otherwise
the logarithmic density gradient become positive. In that cathe matter is
accumulated at $r_{max} $ and the condition
of small optical depth at large $r$ is not fulfilled.
For illustration of a possible solution
consider the intermediate case
$\a = 1$.
The characteristic directions, obtained from Eq. A11, are

$$
{d {\rm y} \over d {\rm x}} =
{1 \over 10}\times (3 \pm \sqrt {129}) \pt
\eqno(A12)
$$

As is known,
the solutions for $w_i$ are unique in the region bounded by these
curves, and
we see that $ (d {\rm y}/d {\rm x})_+ > 0$ while
$ (d {\rm y}/ d{\rm x})_- < 0$,
namely the domain of influence increases (in $\nu$) as
$r$ increases. It is known that a blackbody spectrum
of a moving medium propagating with velocity $\beta$ and direction $\theta$
is preserved, but an observer at rest
will see a Doppler shifted temperature
$T_{obs}  = \bigr( \g [1 - \beta{\rm cos} (\theta )]\bigl)^{-1}
T_{com}$. However, in our case
we see that even if
the energy density at $r_{min}$
$w_0 \propto \nu^3 /[exp(-\nu/T) - 1]$ and the net flux density vanishes
$ w_1 = 0$,
the spectrum is broadened, and thus deviates from the Planck curve.
This confirms the conclusions of ZSP91
for the case of a monoenergetic beam of electrons.
If we continue our approximation to the optically thin limit
where $w_1 \to w_0  \equiv w$
we obtain from combining Eqs. A6 and A7
a power law spectrum
$$
w(\nu, r_{max})\propto \nu ^{-1.34} \pt
\eqno(A13)
$$
This specific result shows how, under certain condition,
the energy density
converts from a blackbothe transition from $\tau \gg 1$ to $ \tau \ll 1$. Such
a transition zone
can be important for explaining GRB spectra.

\end{document}